\documentclass[prb,twocolumn,showpacs]{revtex4}
\usepackage{graphicx}
\begin{document}

\title{
   Effect of free carriers and impurities 
   on density of states and optical spectra \\ 
   of two-dimensional magneto-excitons}

\author{
   Anna G{\l}adysiewicz, Leszek Bryja, and Arkadiusz W\'ojs}
   
\affiliation{\mbox{
   Institute of Physics,
   Wroc{\l}aw University of Technology, 
   Wybrze\.ze Wyspia\'nskiego 27, 50-370 Wroclaw, Poland}}

\author{
   Marek Potemski}

\affiliation{
   Grenoble High Magnetic Field Laboratory, CNRS,
   F-38042 Grenoble Cedex 9, France}

\begin{abstract}
Density of states (DOS) and absorption spectrum of weakly doped, 
narrow quantum wells in high magnetic fields are calculated by 
realistic exact diagonalization.
The systems containing an electron--hole pair with and without
an additional, second electron are compared.
In DOS, the exciton--electron interaction is shown to fill the
gaps between Landau levels and to yield additional discrete peaks
corresponding to bound trion states.
In absorption, interaction with the additional free electron causes 
no shift or renormalization of main, excitonic peaks.
However, it results in additional, weaker peaks associated with
bound trions in the lowest or higher Landau levels.
The calculation is supplemented with experimental photoluminescence 
and photoluminescence-excitation studies of two-dimensional holes 
and electrons in high magnetic fields.
\end{abstract}
\maketitle

\section{Introduction}

Trions (also called charged excitons) are bound states of a neutral 
exciton (electron--hole pair, $X=e+h$) with an additional carrier,
either electron or valence hole, for a negative ($X^-=2e+h$) or 
positive trion ($X^+=2h+e$), respectively.\cite{Lampert58,Kheng93}
Neutral and charged excitons occur naturally in photoluminescence 
(PL) experiments, in which creation or annihilation of $e$--$h$ 
pairs accompanies interband absorption or emission of light.
\cite{Bassani75}
While excitons are expected in charge-neutral systems, formation 
of trions depends on the presence of free carriers.

The trion binding energy $\Delta$ is defined as the effective 
attraction between an exciton and an additional carrier ($\Delta
=E[X]+E[e]-E[X^\pm]$, where $E[\dots]$ is the ground energy of 
a given complex).
An important material parameter affecting $\Delta$ is the ratio 
of electron and hole effective masses,\cite{Narvaez01} $\eta=
\mu_e/\mu_h$; for $\eta=0$ the $X^-$ dynamics reduces to a familiar
problem of the $D^-$ center.\cite{Huant90,Dzyubenko02}
Originating from charge-dipole interaction, trion binding is 
usually much weaker than the exciton binding.
However, its strong enhancement by spatial confinement and/or 
magnetic field $B$ was predicted.\cite{Stebe89}
The pioneering experiment of Kheng {\em et al.}\cite{Kheng93} 
in CdTe, followed by a series of measurements in GaAs
\cite{Buhmann95,Finkelstein95,Shields95a,Gekhtman96} and ZnSe
\cite{Astakhov99,Homburg00} confirmed that it is sufficient for 
the trion's detection.
The $X^+$ was also observed\cite{Shields95b,Glasberg99} and shown 
to be different from $X^-$ (due to the $\eta\leftrightarrow\eta^{-1}$ 
asymmetry).

In a typical experimental configuration, a quantum well containing 
a quasi-two-dimensional (2D) electron (or hole) gas is placed in 
the perpendicular field.
The field should be sufficiently strong to induce magnetic 
quantization of single-particle states into macroscopically 
degenerate Landau levels (LLs), with the inter-LL (cyclotron) 
spacing $\hbar\omega_c$ which is at least comparable to the 
effective excitonic Rydberg ($Ry\approx5.5$~meV in GaAs).
The well width $w$ should not be much greater than the effective 
excitonic Bohr radius ($a_B\approx10$~nm in GaAs) and the magnetic 
length ($\lambda=\sqrt{hc/eB}\approx8.1$~nm at $B=10$~T).

The quantum-mechanical problem of a 2D exciton in a high magnetic 
field goes back several decades.\cite{Gorkov67,Bychkov81,Kallin84}
The continuous (owing to the neutral charge) energy dispersion
in some idealized situations is known analytically, and the
deviations due to inter-LL mixing or finite well width have been
studied experimentally and numerically.\cite{Lozovik02}
The simple $q=0$ optical selection rule ($q$ being the wavevector 
of an annihilated $e$--$h$ pair) results from the small momentum
carried by the absorbed/emitted photon.
In experiment, the excitonic recombination is usually observed
in PL of quantum wells containing no free carriers.

2D trions at high $B$ have also been extensively studied in 
the past.\cite{BarJoseph05}
Different theoretical/computational approaches and the key results 
have been discussed in an exhaustive review by Peeters, Riva, and
Varga.\cite{Peeters01}
Depending on the parameters (e.g., $w$ and $B$), the trion energy 
spectrum contains one or more bound states, which can be conveniently 
labeled by total spin $S$ of the pair of electrons and total angular 
momentum $M$.
At small $B$, the only bound $X^-$ state is the spin-singlet 
with $S=0$ and $M=0$, equivalent to a 2D Hydrogen ion.
\cite{Kheng93,Buhmann95,Finkelstein95,Shields95a,Gekhtman96}
In the (unrealistic) limit of very high $B$ and vanishing $w$, 
the singlet unbinds, and it is replaced by the spin-triplet 
with $S=1$ and $M=-1$.\cite{Wojs95,Palacios96}
A triplet trion was identified by Shields {\em et al.}
\cite{Shields95a} (although now it is not clear if it was the 
$M=-1$ triplet ground state).
The singlet--triplet crossover was predicted\cite{Whittaker97} 
in rather high fields (e.g.\ $B\approx30$~T in narrow symmetric 
GaAs wells).
Despite initial difficulties\cite{Hayne99} it was also eventually 
confirmed in several experiments.
\cite{Munteanu00,Yusa01,Vanhoucke01,Astakhov05}
Additional bound states were predicted\cite{Wojs00} at intermediate 
$B$, but (for realistic parameters) they are always less strongly 
bound than the above two.
These states were also confirmed by both independent calculations
\cite{Riva00} and by experiments.
\cite{Yusa01,Vanhoucke01,Andronikov05}

The pair of optical selection rules results from invariance
under (magnetic) translations\cite{Avron78,Dzyubenko00} 
and from the particle--hole symmetry between conduction electrons 
and valence holes.\cite{Lerner81,Dzyubenko83,Macdonald90}
Both these ``geometric'' and ``hidden'' symmetries are at least 
weakly broken in realistic conditions.
Nevertheless, recombination of trions with $M\ne0$ requires
a symmetry-breaking collision (with an impurity or another 
carrier), and therefore, it became customary to label different 
trions as ``bright'' and ``dark'' (meaning having $M=0$ and 
$M\ne0$, respectively).
In most recent experiments,\cite{Hayne99,Yusa01,Andronikov05}
up to three trion states are observed: ``bright'' singlet 
$X^-_{\rm s}$, ``dark'' triplet $X^-_{\rm td}$, and a weakly 
bound ``bright'' triplet $X^-_{\rm tb}$ with $S=1$ and $M=0$.
Vanishing oscillator strength,\cite{Palacios96,Dzyubenko00} 
of $X^-_{\rm td}$ confirmed directly in optical absorption,
\cite{Schuller02} makes it more difficult to observe (even 
in PL) than the other trions.
In fact, this state had not been conclusively identified until 
the work of Yusa {\em et al.},\cite{Yusa01} motivated by earlier
high-field experiments (especially of Hayne {\em et al.},
\cite{Hayne99} showing apparent discrepancy with prediction
\cite{Whittaker97} of singlet--triplet crossing, but also 
several others\cite{Munteanu00} showing inconsistent features 
in trion spectra) and a numerical prediction\cite{Wojs00} 
of an excited triplet state, $X^-_{\rm tb}$.
Most recently, additional weakly bound states were reported
\cite{Astakhov05} in CdTe ($Ry\sim10$~meV, about twice larger 
than in GaAs).

Involving only three particles, trion's quantum dynamics might 
appear to be relatively simple, both conceptually and computationally.
Addition of high magnetic field and confinement does not add much 
complexity, and indeed, quantum numbers and symmetries of all bound
trions (in wide range of realistic conditions) are by now understood.
In short, the dynamics depends on competition of several energy scales, 
with the following characteristic values for our example of a 15~nm 
symmetric GaAs well at $B=10$~T: cyclotron gap ($\hbar\omega_c
\sim18$~meV for electrons and $\sim4$~meV for heavy holes), Coulomb 
energy ($e^2/\lambda\sim14$~meV), and a small Zeeman gap.
The well width $w$ comes in two places, defining excitation gaps 
to higher subbands (gap to the second subband is $\sim50$~meV for 
electrons and $\sim10$~meV for the holes) and affecting in-plane 
Coulomb matrix elements (this effect is parametrized by $w/\lambda\sim2$).
Also, even weak asymmetry between electron and hole subband wavefunctions 
$\chi_s(z)$ can considerably affect the trion binding, since it leads 
to different magnitudes of $e$--$e$ and $e$--$h$ interactions, which 
no longer cancel in $\Delta$ (this effect is essential in asymmetric 
wells, not considered here, where the asymmetry depends on carrier 
concentration and/or additional gates).

However, the above optimistic (and popular) view hides the fact that 
good understanding of the role of trions in PL experiments must include 
variety of complications due to coupling to the environment (e.g., 
complex single-particle energy band structure with nonparabolic and 
anisotropic dispersions, interaction with free carriers, lattice 
defects, and phonons, spin-orbit effects, etc.).
Some quantities (e.g., the binding energies, especially of the triplet 
states) can be calculated rather accurately.\cite{Wojs00,Riva00} 
However, others (such as the critical values of $B$ and $w$ for the 
singlet--triplet transitions) depend so sensitively on the (often 
unknown) parameters that their quantitative modeling turns out 
somewhat pointless.
Another unsolved (quantitatively) and important problem is the 
kinetics of trions,\cite{Jeukens02,Plochocka04} involving their 
binding/unbinding ($X^-\leftrightarrow X+e^-$) and, at high $B$, 
orbital/spin relaxation ($X^-_{\rm s}\leftrightarrow X^-_{\rm td}
\leftrightarrow X^-_{\rm tb}$).

In this article, we analyze the effect of trions on the optical 
absorption spectrum of 2D electrons in a magnetic field.
Thus, in addition to the earlier studied bound trion states,
\cite{Wojs00,Riva00} the entire low-energy $2e+h$ spectrum is 
considered. 
In a somewhat related work, asymmetry of trion absorption peaks 
at small $B$ was discussed by Stebe {\em et al.}\cite{Stebe98}
The inclusion of only two electrons and one hole in the model 
restricts it to the ``dilute'' regime, defined by a small value 
of the filling factor, $\nu\ll1$ ($\nu$ is defined as the number 
of electrons $N$ divided by the LL degeneracy $g$; alternatively, 
$\nu=2\pi\varrho\lambda^2$ where $\varrho$ is the 2D electron 
concentration).
It is remarkable that in some systems the immersed trion is only 
weakly (perturbatively) affected by the surrounding electrons.
In narrow wells and in high magnetic fields, this occurs at 
$\nu<{1\over3}$ due to ``Laughlin correlations''\cite{Laughlin83} 
between electrons and trions,\cite{Wojs99} preventing strong 
$e$--$X^-$ collisions (indeed, in wider wells trions seem to be 
strongly coupled to the electrons and cannot be regarded as simple 
three-body quasiparticles\cite{Sanvitto02}).

The density of states (DOS) and absorption spectra are calculated 
numerically by exact diagonalization in Haldane's spherical geometry.
\cite{Wu76,Haldane83}
The figures were drawn for a particular choice of a symmetric 
$w=15$~nm GaAs quantum well at $B=10$~T.
Since absorption into bound trion states was studied earlier, 
we concentrate on the effects of the interaction of the exciton 
with a (single) {\em unbound} electron.
The main conclusions are redistribution of the density of states 
(by filling the gaps between the LLs) and the emergence of additional, 
trion peaks in the absorption spectrum (also in the excited LLs).

The calculation is supplemented with the results of experimental 
polarization-resolved photoluminescence (PL) and 
photoluminescence-excitation (PLE) studies of a 2D hole and electron 
gases in a symmetric 15~nm GaAs well.
Presented spectra reveal absorption into a pair of bright trions 
in the lowest LL and emission from these two states, the exciton, 
and the dark triplet trion.

\section{Model}

The $e+h$ and $2e+h$ energy spectra are calculated by exact
diagonalization of the hamiltonian matrix on a Haldane sphere, 
convenient for modeling an infinite plane with 2D translational 
invariance.
In this geometry, the magnetic field normal to the spherical 
surface of radius $R$ is produced by a Dirac monopole od strength 
$2Q=4\pi R^2/\phi_0$ ($\phi_0=hc/e$ is the magnetic flux quantum).
Through the following relation, $R^2=Q\lambda^2$, monopole strength
determines surface curvature (in the magnetic units).

The single-particle states are called monopole harmonics.\cite{Wu76}
They are the eigenstates of angular momentum $l$ and its projection 
$m$ on the $z$-axis. 
The lowest shell, corresponding to the lowest LL, has $l=Q$ and 
finite degeneracy $g=2Q+1$.
Higher shells, corresponding to the excited LLs labeled by index $n$, 
have $l=Q+n$.

The energy of the $n$th shell is $\varepsilon_n=\hbar\omega_c(n+
{1\over2})+\hbar^2n(n+1)/2\mu R^2$.
It contains the cyclotron gap (with $\omega_c=eB/\mu c$, where $\mu$ 
is the effective mass) and an additional, curvature-dependent term.
In order to model a real structure, we take advantage of the LL 
structure of monopole harmonics, but replace $\varepsilon_n$ with
the correct energies of electron and hole LLs.
E.g., at $B=10$~T, we use $\hbar\omega_c=17.8$~meV for electrons 
and 3.7~meV for the heavy hole (the latter value taken after Cole 
{\em et al.} \cite{Cole97} for $w=15$~nm).

The second-quantization hamiltonian reads
\begin{equation}
   H = \sum_i c_i^\dagger c_i \varepsilon_i
     + \sum_{ijkl} c_i^\dagger c_j^\dagger c_k c_l v_{ijkl}.
\label{eqHam}
\end{equation}
An additional term $\sum_{ij} c_i^\dagger c_j u_{ij}$ will also be 
included to describe interaction with a positive or negative point
charge.
Here, $c_i^\dagger$ and $c_i$ are operators creating and annihilating 
a conduction electron or a valence hole, in the state labeled by a 
composite index $i$ containing all relevant single-particle quantum 
numbers (band $\beta$, subband $s$, LL index $n$, angular momentum 
$m$, and spin $\sigma$).

The Coulomb interaction matrix elements $v$ and $u$ in the basis of 
monopole harmonics are known analytically for all particles confined 
to a 2D surface of the sphere.
However, to account for a finite width of the real quantum well, 
we have integrated all matrix elements numerically, in 3D, using 
form-factors appropriate for the actual electron and hole subband 
wavefunctions $\chi_s(z)$.
Mixing with higher quantum well subbands (excited states in the 
$z$-direction, labeled by $s>0$) is not very strong in a narrow and 
symmetric well due to high quantization energy (for $w=15$~nm and
Al$_0.35$Ga$_{0.65}$As barriers, it is 49.1~meV and 11.5~meV to the 
first excited electron and hole subbands, respectively; calculation 
after Ref.~\onlinecite{Tan90}) and the parity conservation.
Nevertheless, it is not quite negligible for the bound states
(states with strong interactions).

Without impurities, the $e+h$ and $2e+h$ eigenstates of $H$ are 
labeled by total angular momentum $L$ and its projection $L_z$.
When converting these quantities to the planar geometry, neutral 
and charged states must be treated differently:
$L$ of an $e$--$h$ pair represents wavevector $q=L/R$, and for a 
$2e+h$ state it corresponds to $M=L-Q$.
The $2e+h$ eigenstates are also labeled by spin $S$ of the pair
of electrons and its projection $S_z$.
The calculation need only be performed in the $L_z=S_z=0$ subspace
and the appropriate Zeeman shift can be added at the end to the 
energies of each triplet ($S=1$) state.

With an impurity placed at a north pole of the sphere, $L_z$ is still 
conserved, but $L$ is not.
Only the $S_z=0$ subspace need be considered, but a separate 
diagonalization must be performed for each $L_z$.

The $2e+h$ diagonalization was carried out in configuration-interaction
basis, $\left|i,j;k\right>=c_i^\dagger c_j^\dagger c_k^\dagger
\left|{\rm vac}\right>$.
Here indices $i$ and $j$ denote the occupied electron states, $k$ 
describes the hole, and $\left|{\rm vac}\right>$ is the vacuum state.
Similarly, the basis for the $e+h$ calculation was $\left|i;k\right>$.
As mentioned earlier, only the spin-unpolarized states with $L_z\equiv 
m_i+m_j-m_k=0$ are included in the basis.
To find eigenstates corresponding to given $(L,S)$ we used a modified 
Lanczos algorithm, with additional projection of Lanczos vectors at 
each iteration.

\section{Results and Discussion}

\subsection{Density of States}

\begin{figure}
\includegraphics[width=3.2in]{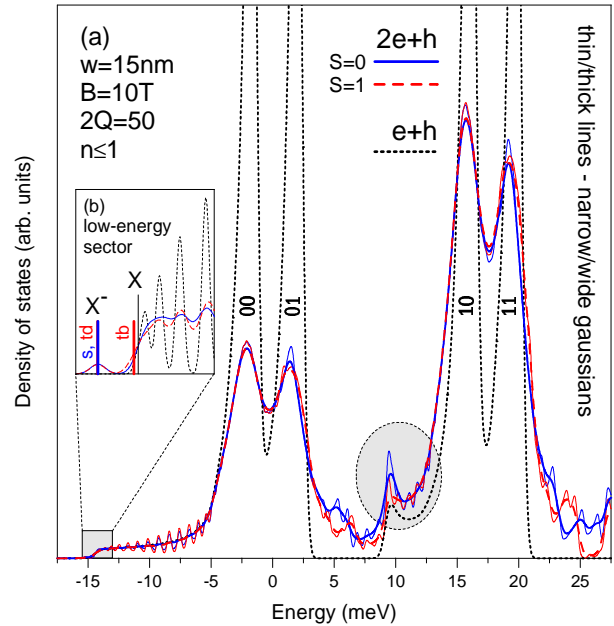}
\caption{(color online) 
   Density of states (DOS) of $e+h$ and $2e+h$ systems 
   in a symmetric GaAs quantum well of width $w=15$~nm 
   at magnetic field $B=10$~T,
   calculated on Haldane sphere 
   with a large magnetic monopole strength $2Q=50$
   including only two lowest electron and hole LLs 
   ($n\le1$).
   Continuous curves were obtained from finite-size 
   discrete energy spectra by gaussian broadening.
   For $2e+h$, DOS is divided by $g=2Q+1$ (LL degeneracy 
   of the second electron), and singlet and triplet 
   configurations (spin $S=0$ and 1) are drawn separately.
   For $e+h$, LL indices ``$n_en_h$'' mark the strongest peaks.
   Inset: magnified lowest-energy sector (vertical lines 
   mark position of discrete bound trions and the excitonic 
   ground state.}
\label{fig1}
\end{figure}

We begin by the calculation of the density of states ${\rm DOS}=
d\Gamma/dE$ (the number of states $\Gamma$ per unit of energy $E$).
In Fig.~\ref{fig1} we compare the results for $e+h$ and $2e+h$
obtained for a rather large $2Q=50$ and including only the lowest 
two electron and hole LLs ($n\le1$).
The discrete energy spectra obtained from finite-size calculation 
were converted to the continuous lines shown in the figure by
broadening with Gaussians,
\begin{equation}
   {\rm DOS}(E)=\sum_i \mathcal{G}_\delta(E-E_i),
\label{eqDOS}
\end{equation}
where the summation goes over all energy levels $E_i$ and 
$\mathcal{G}_\delta(x)=\delta^{-1}\pi^{-1/2}\exp(-x^2/\delta^2)$.
Thin and thick lines correspond to two different broadening widths 
$\delta=0.2$ and 0.5~meV. 
Blue solid, red dashed, and black dotted lines were used to
distinguish $2e+h$ ($S=0$ and 1 plotted separately; only $S_z=0$ 
is shown for $S=1$) from $e+h$.
To compare DOS for $e+h$ and $2e+h$, the latter curve was rescaled 
from the value defined by Eq.~(\ref{eqDOS}) by $g^{-1}=(2Q+1)^{-1}$ 
(i.e., divided by the number of states available to the second electron).
Energy $E$ is measured from the noninteracting configurations 
in the lowest LL.

For a noninteracting $e$--$h$ pair, DOS consists of discrete peaks
labeled by the LL indices for both particles, ``$n_en_h$''.
The black dotted curve demonstrate the effect of $e$--$h$ interaction.
Within each pair of LLs, magneto-excitonic dispersion becomes flat 
at long wavevectors $q$, corresponding to the vanishing $e$--$h$ 
attraction.
Therefore, although smeared toward lower energies, strong ``$n_en_h$'' 
peaks persist.
In our calculation, restricted area of the sphere prohibits the pair 
to separate in the $q\rightarrow\infty$ limit (the range of $q=L/R$ 
is restricted by $L\le2Q$), and the ``$n_en_h$'' peaks have finite 
height and are displaced to the left by the remnant attraction 
($\sim e^2/2R=(2\sqrt{Q})^{-1}\,e^2/\lambda$).
This redshift ($\sim1.4$~meV for $2Q=50$) is a finite-size artifact.
The low-energy tail extends from each ``$n_en_h$'' peak over the range
of the excitonic binding energy within the corresponding pair of LLs.
Since $e^2/\lambda$ is larger than $\hbar\omega_c$ of the holes, these 
tails essentially close the gaps between the neighboring hole LLs.
Thin lines (shown in the magnified low-energy sector in the inset) 
reveal artificial size quantization on a sphere ($L=0$, 1, 2, \dots).
On the other hand, an interesting {\em real} feature is the maximum 
at the beginning of the ``$10$'' tail ($E\approx10$~meV).
It is due to the fact that excitonic dispersion for $n_e\ne n_h$ is
nonmonotonic at small wavevectors $q$ (the ground state occurs at 
$q>0$, leading to $dE/dq=0$ and $d\Gamma/dq\propto q>0$, i.e., to 
a singularity in $d\Gamma/dE$).

The curves for $2e+h$ are also dominated by the noninteracting peaks 
``$n_en_e'n_h$'' corresponding to three unbound particles confined to 
different combinations of LLs. 
Low-energy tails describe the exciton with an additional unbound electron,
(note oscillations due to excitonic size quantization).
Bound trion states appear as discrete peaks below the continuous tails,
well visible only in the inset, additionally marked with red and blue 
bars.
All three trions: $X^-_{\rm s}$, $X^-_{\rm td}$, and $X^-_{\rm tb}$
appear bound (excitonic ground state is marked by a black bar for 
comparison).
However, weak ($<1$~meV) binding of $X^-_{\rm s}$, virtually equal to 
$X^-_{\rm td}$, is an artifact of the unrealistic $n\le1$ restriction.

The emergence of bound trion states below the exciton's continuum 
is one noticeable difference between the $e+h$ and $2e+h$ DOS.
Another significant effect of the exciton--electron interaction 
is further (compared to one due to excitonic $e$--$h$ interaction)
smearing of the LLs, i.e., transfer of the density of states away 
from LLs and filling the gaps between them.
Also, DOS within the inter-LL regions shows features related to the 
interactions in a nontrivial manner.
For example, spin dependence of structures at $E\approx5$~meV and 
$E\approx25$~meV reveals their connection with the (obviously, 
spin-sensitive) exciton--electron interaction.
Note that the $e+h$ peak at $E\approx10$~meV, identified earlier 
with an inter-LL exciton, persists in the $2e+h$ curves regardless
of spin configuration (to confirm its {\em one}-electron nature).

The $n\le1$ restriction to only two lowest LLs was helpful in 
demonstrating LL smearing and emergence of additional peaks between 
LLs due to $e$--$h$ and $X$--$e$ interactions used in Figs.~1 and 2.
However, it gives incorrect exciton and trion energies and, more 
importantly, ignores the contribution to DOS coming from the 
neglected higher hole LLs (recall that hole's $\hbar\omega_c$ is 
only 3.7~meV at $B=10$~T, much smaller than electron's 17.8~meV).

\begin{figure}
\includegraphics[width=3.2in]{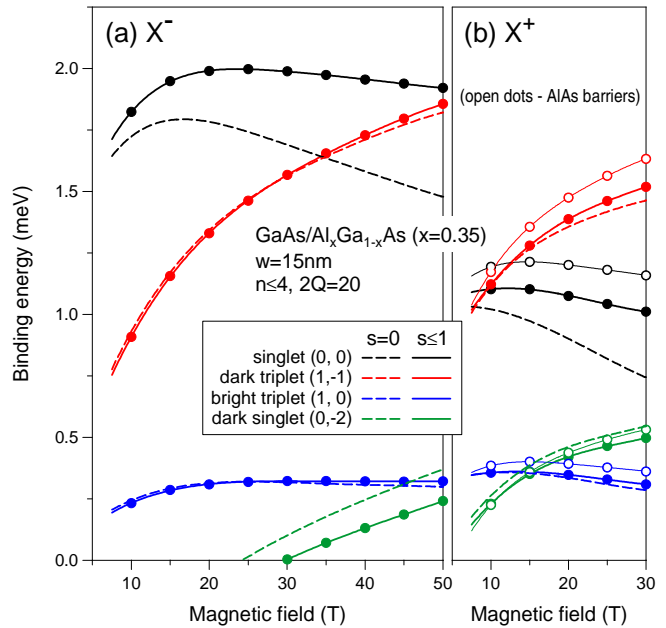}
\caption{(color online) 
   Binding energies $\Delta$ of four different negative (a) and
   positive (b) trions $X^\pm$ (the pair of labels in parentheses 
   are two-electron spin $S$ and total angular momentum $M$) in 
   a symmetric GaAs quantum well of width $w=15$~nm as a function 
   of magnetic field $B$, calculated on Haldane sphere with a 
   magnetic monopole strength $2Q=20$ including five lowest LLs 
   ($n\le4$) and either one ($s=0$) or two lowest subbands ($s\le1$) 
   for each electron and hole.
   Continuous curves were obtained by interpolation from 
   exact-diagonalization at the values of $B$ marked with dots.}
\label{fig2}
\end{figure}

More accurate trion binding energies (for both $X^-$ and $X^+$)
are plotted in Fig.~\ref{fig2} as a function of $B$.
These values were obtained by including five LLs ($n\le4$) and 
up to two subbands ($s\le1$) for each electron and hole.
The lowest-subband ($s=0$) calculation is similar to 
Ref.~\onlinecite{Wojs00}, but it used more accurate Coulomb 
matrix elements, calculated for the actual $\chi_0(z)$.
For $X^-$, an additional, weakly bound ``dark singlet'' state 
occurs only at quite high fields.
The curves for $s\le1$ demonstrate that in a narrow symmetric well 
the subband mixing is most efficient for the $M=0$ singlet, while 
the lowest-subband approximation is accurate for both triplets.

In our calculations we assumed the Al fraction of $x=0.35$ in the
Al$_x$Ga$_{1-x}$As barriers.
This restriction is justified by the fact that the binding energies 
of $X^-$ are almost insensitive to the increase of Al fraction all
the way up to $x=1$.
The only exception is data for $X^+$ marked with open dots in 
Fig.~\ref{fig2}(b), obtained for pure AlAs barriers.
Evidently, the $X^+$ binding energies depend more strongly on the 
barrier height, which must be taken into account in a realistic 
model.

\begin{figure}
\includegraphics[width=3.2in]{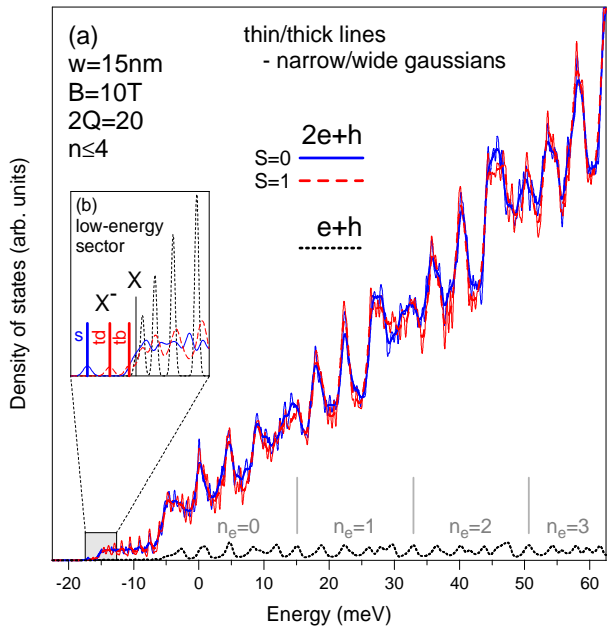}
\caption{(color online) 
   The same as in Fig.~\ref{fig1} but for $2Q=20$, $n\le4$, and
   without dividing $2e+h$ DOS by $g=2Q+1$.}
\label{fig3}
\end{figure}

More accurate DOS is shown in Fig.~\ref{fig3}, including five LLs 
($n\leq4$) but still only the lowest subband.
Since dimension of the Hilbert space quickly grows with $n_{\rm 
max}$ and $2Q$, we were forced to use a smaller $2Q=20$ in this 
case (for $2e+h$ this yields dimension 58,875 for the $L_z=S_z=0$ 
subspace, whose {\em all} eigenenergies must be calculated to
plot DOS).
Especially at higher energies, the curves would become very 
complicated due to an increasing number of overlapping peaks 
corresponding to different combinations of $n_e$ and $n_h$
-- if not only 5, but {\em all} LLs were included for the hole.
By counting electron cyclotron gaps from the lowest LL peak at
$E\approx-e^2/2R$, energies corresponding to consecutive $n_e$'s
have been marked with gray vertical lines.
Comparison of the curves for $e+h$ (here plotted without rescaling 
by $g^{-1}$) and $2e+h$ shows that the effect of smearing the LLs 
and filling the gaps between them due to $X$--$e$ interaction is 
only enhanced when more LLs are included.
In the inset, the binding energies of all trion peaks are already 
well converged for $n\le4$ (note however that the lowest subband 
approximation considerably affects especially the singlet state);
the exact values are $\Delta=1.72$, 0.93, and 0.24~meV for 
$X^-_{\rm s}$, $X^-_{\rm td}$, and $X^-_{\rm tb}$, respectively.

\begin{figure}
\includegraphics[width=3.2in]{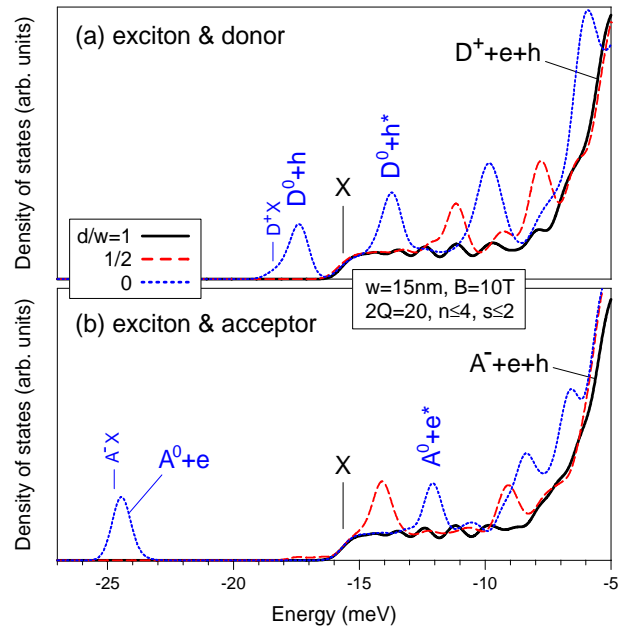}
\caption{(color online) 
   Density of states (DOS) of $e+h$ in a symmetric GaAs quantum 
   well of width $w=15$~nm at magnetic field $B=10$~T, in the 
   presence of a positive (a) and negative (b) impurity at
   different distances $d$ from the center of the well, calculated 
   on Haldane sphere with magnetic monopole strength $2Q=20$
   including five LLs ($n\le4$) and three subbands ($s\le2$) 
   for electron and hole.
   Continuous curves were obtained by gaussian broadening.
   Marked bound states: 
   $D^0=D^++e$,     $D^+X=D^++X$, 
   $A^0=A^-+h$, and $A^-X=A^-+X$; 
   $e^*$ and $h^*$ denote electron and hole 
   in the excited ($n=1$) LL.}
\label{fig4}
\end{figure}

An ionized impurity can have a similar effect on the $e+h$ DOS 
to that of an additional electron.
This impurity can be either a positive or negative point charge
(ionized donor $D^+$ or ionized acceptor $A^-$) placed at 
a distance $d$ away from the center of the quantum well.
The effects of $D^+$ and $A^-$ are not equivalent due to the 
$e$--$h$ asymmetry. 
Two frames of Fig.~\ref{fig4} show how the $e+h$ DOS changes 
in the presence of $D^+$ or $A^-$ placed at $d/w={1\over2}$ 
and 0 (edge and center of the well).
In plotted energy range (up to the first excitonic maximum 
``00''), the curves for $d/w=1$ are almost identical to those 
without an impurity).
One or more bound $D^+X$ or $A^-X$ states (analogous to trions)
emerge below the excitonic tail, at the position sensitive to $d$.
Being localized (and nondegenerate), they do not contribute to the 
macroscopic DOS.

When the impurity approaches the well, a strong peak detaches 
from ``00'' and moves to the left through the excitonic tail.
It corresponds to the (macroscopically degenerate) $D^0+h$ 
or $A^0+e$ configuration with the unbound $e$--$h$ pair.
In our example, it passes the tail's edge ($X$ ground state) when
the impurity is already inside the well, at $d\approx3$~nm for 
$D^+$ and 5.5~nm for $A^-$.
For $d=0$, especially the $A^0$ is bound much more strongly than $X$.
Thus, the strongest effect of an impurity is that, depending on $d$, 
it can bring down macroscopic DOS corresponding to unbound $e$--$h$ 
below the free excitonic ground state.
Certainly, the localized $D^+X$ and $A^-X$ states involving the 
$e$--$h$ binding still remain the absolute ground states.
Here, the qualitative difference caused by the impurity is that DOS 
raises abruptly from essentially zero at the nondegenerate bound state 
to the continuum of degenerate unbound states (instead of through an 
excitonic tail).
Especially the marginally bound $A^-X$ is hardly distinguished from
the continuum.

For $d=0$ we have also marked the $D^0+h^*$ and $A^0+e^*$ peaks 
corresponding to the unbound hole or electron in a higher LL,
and thus originating from the ``01'' and ``10'' peaks without 
an impurity.
These peaks are separated from $D^0+h$ and $A^0+e$ by a cyclotron
gap; thus, an impurity mixes the LL peak structure with an excitonic
tail due to $e$--$h$ interaction.

\subsection{Oscillator Strength}

Let us now turn to the calculation of oscillator strength $\Omega$ 
for the ${\rm vac}\leftrightarrow e+h$ and $e\leftrightarrow2e+h$ 
optical transitions (``$\rightarrow$'' for absorption; ``$\leftarrow$''
for emission).
For a pair of initial and final states, e.g., $\psi=e$ and 
$\phi=2e+h$, it is calculated from Fermi's golden rule,
\begin{equation}
   \Omega_{\psi\phi}=\sum_k
   \left|
   \left<\phi\right|c_k^\dagger c_{\bar k}^\dagger\left|\psi\right>
   \right|^2.
\end{equation}
Here, $k=[\beta,n,m,\sigma]$ and $\bar k=[\bar\beta,n,-m,-\sigma]$, 
and the summation runs over all electron states $k$ in the conduction 
band and all corresponding hole states $\bar k$ in the valence band.
Note that according to convention of Eq.~(\ref{eqHam}), $c_k^\dagger$ 
creates electrons or holes, depending on band index $\beta$, leading 
to the reversed sign of $m$ and $\sigma$ in $\bar k$.
The oscillator strength for the recombination of initial $\phi=e+h$ 
or $2e+h$ states, in the latter case summed over all final $\psi=e$ 
states, can be expressed as a function of the initial energy $E=E_\phi$,
\begin{equation}
   \Omega(E)=\sum_{\psi\phi}\Omega_{\psi\phi}\,\delta(E_\phi-E),
\label{eqOm1}
\end{equation}
This is the $e+h$ or $2e+h$ ``optical density of states'' (ODOS).
Alternatively, oscillator strength (if necessary, weighted by the 
occupation function $\Theta_\psi$ for the initial states $\psi=e$)
can be expressed as a function of the photon energy ${\cal E}=E_\phi
-E_\psi$,
\begin{equation}
   \Omega({\cal E})=\sum_{\psi\phi}\Omega_{\psi\phi}\Theta_\psi\,
   \delta(E_\phi-E_\psi-{\cal E}).
\label{eqOm2}
\end{equation}
This is the $\psi\rightarrow\phi$ absorption spectrum (equivalent 
to $\phi$'s ODOS for $\psi={\rm vac}$, but not for $\psi=e$).

\begin{figure}
\includegraphics[width=3.2in]{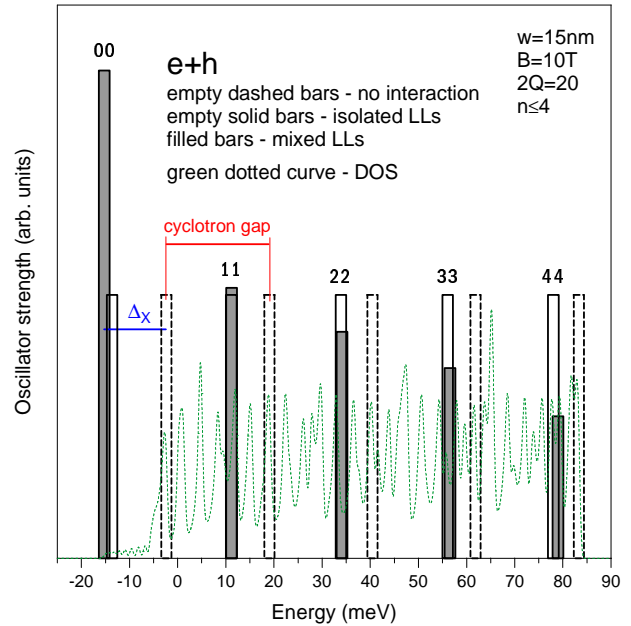}
\caption{(color online) 
   Optical density of states (ODOS) of $e+h$, 
   equivalent to absorption spectrum of an exciton, 
   in a symmetric GaAs quantum well of width $w=15$~nm 
   at magnetic field $B=10$~T,
   calculated on Haldane sphere 
   with a magnetic monopole strength $2Q=20$
   including five lowest electron and hole LLs ($n\le4$).
   Results without $e$--$h$ interaction, with interaction
   within isolated LLs (peaks marked with LL indices ``$n_en_h$''),
   and with all interaction effects (including both intra- 
   and inter-LL scattering) are shown.
   The $e+h$ DOS from Fig.~\ref{fig3} is drawn for reference.}
\label{fig5}
\end{figure}

In Fig.~\ref{fig5} we plot ODOS of $e+h$.
The parameters ($w=15$~nm and $B=10$~T) and Hilbert space ($n\le4$
and $2Q=20$) are the same as in Fig.~\ref{fig3}.
If the Coulomb energy $e^2/\lambda$ were much smaller than the cyclotron 
gaps, then the only optically active states would be the $q=0$ excitons 
with electron and hole confined to the same LLs ($n_e=n_h\equiv n$).
As shown with empty bars, these excitonic peaks ``$nn$'' in $\Omega({\cal 
E})$ would have equal height and occur at $E(n)=n\hbar(\omega_{c,e}+
\omega_{c,h})-\Delta_X(n)$, where $\Delta_X(n)$ is the exciton binding 
energy in the $n$th LL.
These energies coincide with the low-energy edges of the excitonic tails 
of the ``$nn$'' peaks in DOS (replotted with a thin dotted line), where 
degeneracy is absent and DOS vanishes.
Since the exciton binding $\Delta_X(n)$ decreases as a function of $n$,
the separation between consecutive peaks also decreases and it is always 
larger than the cyclotron gap between the corresponding ``$nn$'' maxima 
in DOS.
For $w=15$ and $B=10$~T, distances between peaks ``00'', ``11'', and 
``22'' (calculated excluding LL mixing by setting $v_{ijkl}=0$ unless 
$n_i=n_j=n_k=n_l\equiv n$) are $E_{11}-E_{00}=24.7$~meV and $E_{22}-
E_{11}=22.8$~meV, both considerably larger than $\hbar(\omega_{c,e}+
\omega_{c,h})=21.5$~meV.

In reality, inter-LL scattering mixes $q=0$ states with different $n$.
As shown with filled bars, this causes shifting of the peaks (and a 
further, small increase of the spacing between neighboring peaks) and 
transfer of oscillator strength from higher to lower energy.
Nevertheless, the effect is perturbative and consecutive peaks can
still be labeled by $n$.
For $w=15$ and $B=10$~T, the first two gaps in $\Omega$ (calculated 
including all matrix elements $v_{ijkl}$) increase to $E_{11}-E_{00}
=26.4$~meV and $E_{22}-E_{11}=23.0$~meV.
The relative magnitudes of the lowest three peaks are $\Omega_{11}/
\Omega_{00}=0.55$ and $\Omega_{22}/\Omega_{00}=0.46$.

\begin{figure}
\includegraphics[width=3.2in]{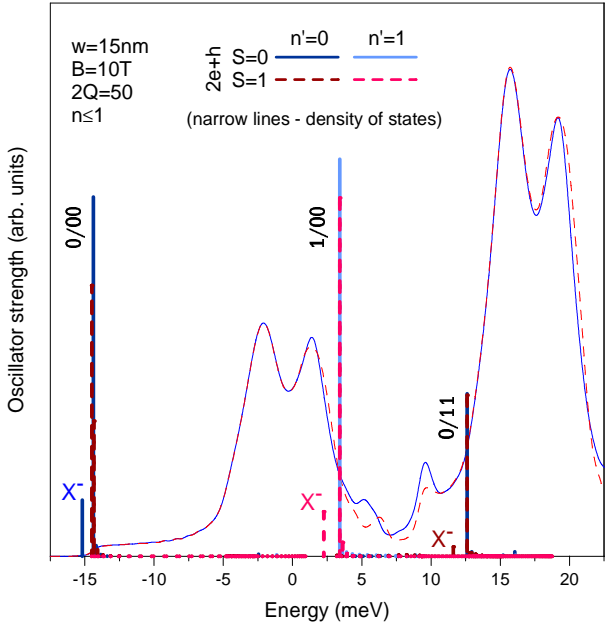}
\caption{(color online) 
   Optical density of states (ODOS) of $2e+h$, 
   in a symmetric GaAs quantum well of width $w=15$~nm 
   at magnetic field $B=10$~T,
   calculated on Haldane sphere 
   with a large magnetic monopole strength $2Q=50$
   including only two lowest electron and hole LLs ($n\le1$).
   Singlet and triplet initial $2e+h$ spin configurations 
   ($S=0$ and 1), and ground and excited final $e$ states 
   ($n'=0$ and 1) are drawn with different lines.
   Strongest peaks are marked by LL indices of the initial
   and final states, ``$n'/n_en_h$'' with $n_e=n_h$.
   Weak peaks associated with bound trions ($X^-$) are also 
   indicated.
   The $2e+h$ DOS from Fig.~\ref{fig1} is drawn for reference.}
\label{fig6}
\end{figure}

In Fig.~\ref{fig6} we plotted ODOS of $2e+h$ calculated using the same 
Hilbert space ($n\le1$ and $2Q=50$) as in Fig.~\ref{fig1}.
Also for this larger system there is no apparent correlation between
the features in DOS and ODOS.
This follows immediately from the $q=0$ selection rule, equivalent to 
the requirement of nondegenerate relative motion of the recombining 
$e$--$h$ pair, preventing high DOS of a $2e+h$ state involving such 
a pair.

In analogy to $e+h$, the main peaks can be labeled by ``$n'/nn$'',
and they correspond to a $q=0$ exciton created/annihilated on the
$n$th LL, in the presence of the second, $\psi$-electron on the
$n'$th LL.
In those main peaks ``$n'/nn$'', the second electron is not bound 
to the created/annihilated exciton; they describe excitonic optical 
processes, weakly affected by the exciton--electron scattering.
Optically active bound trions (with electrons and holes in different
LLs) appear in form of weaker peaks displaced from ``$n'/nn$'' by 
the binding energy.
In the lowest LL, the only well resolved trion is the singlet (dark 
triplet by definition has $\Omega=0$ and bright triplet is too weakly 
bound to be distinguished from exciton in this energy scale).
In the triplet trion at $E\approx2.5$~meV one of the electrons is
in the $n=1$ LL.
Although unstable against inter-LL relaxation, this state can form
by capturing a ``00'' photo-exciton by a thermally excited electron.

Note that ``shake-up'' transitions\cite{Finkelstein96} ($n'/nn
\leftrightarrow n''$ with $n'\ne n''$, i.e., combination of ``$nn$'' 
recombination with $n'\leftrightarrow n''$ cyclotron excitation of 
the second electron) are forbidden\cite{Dzyubenko04} for an isolated 
trion due to invariance under 2D (magnetic) translations.
This selection rule does not preclude replicas of an exciton and an 
{\em unbound} second electron, but these transitions have negligible 
intensity for $\nu\sim g^{-1} \ll1$ and cannot be identified in 
Fig.~\ref{fig6}.

\begin{figure}
\includegraphics[width=3.2in]{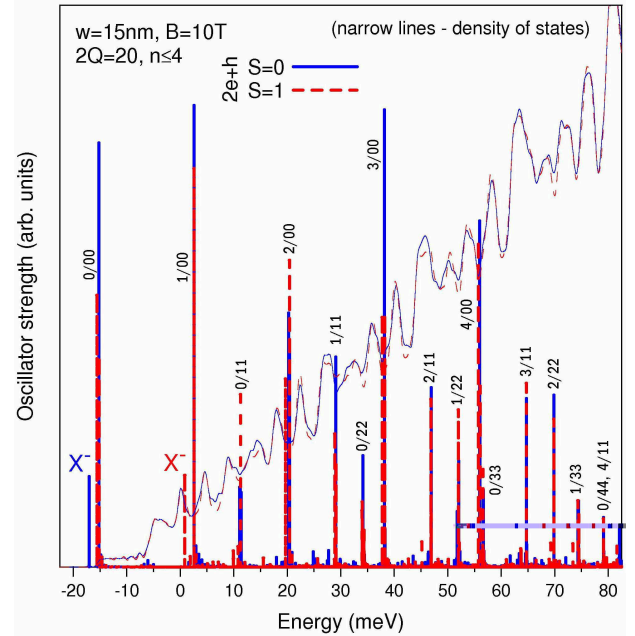}
\caption{(color online) 
   The same as in Fig.~\ref{fig6} but for $2Q=20$ and $n\le4$.
   The strongest peaks are identified as excitonic ``$n'/nn$'' 
   transitions or trion recombination ($X^-$).}
\label{fig7}
\end{figure}

Fig.~\ref{fig7} is similar to Fig.~\ref{fig6}, but it shows $2e+h$ 
ODOS calculated with 5 LLs taken into account ($n\le4$ and $2Q=20$).
The picture becomes fairly complicated, but the idea is the same.
Dominant peaks correspond to excitonic transitions and can be labeled 
by ``$n'/nn$'' (in the calculated spectrum, the assignment is 
straightforward due to angular momentum conservation in the optical 
transition, here causing the $L=Q+n'$ selection rule).
A great number of weaker transitions that involve exciton--electron
interaction emerge around the excitonic peaks.
In particular, bright trions appear below ``$0/00$'' and ``$1/00$''
(singlet and triplet, respectively).

\begin{figure}
\includegraphics[width=3.2in]{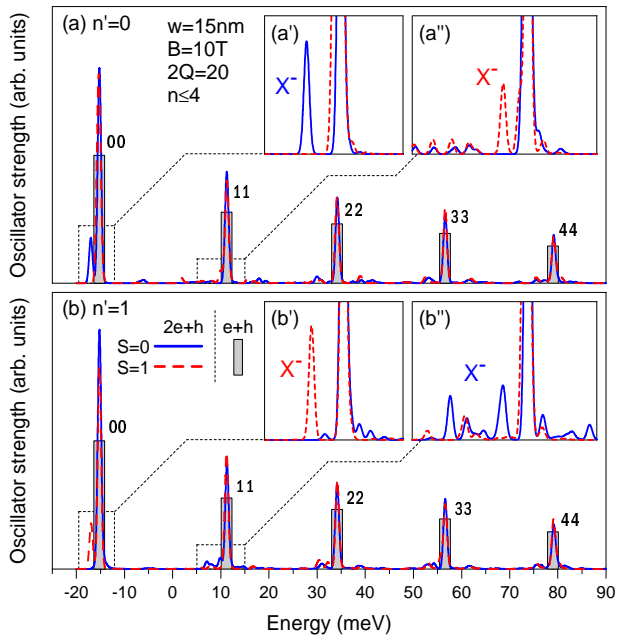}
\caption{(color online) 
   $2e+h$ absorption spectra in a symmetric GaAs quantum well 
   of width $w=15$~nm at magnetic field $B=10$~T,
   calculated on Haldane sphere 
   with a magnetic monopole strength $2Q=20$
   including five lowest electron and hole LLs ($n\le4$).
   Continuous curves were obtained by gaussian broadening.
   Singlet and triplet $2e+h$ spin configurations 
   ($S=0$ and 1) are drawn separately.
   Frames (a) and (b) correspond to the initial-state electron 
   in the lowest or first excited LL ($n'=0$ or 1).
   The strongest peaks marked as ``$nn$'' correspond to excitonic 
   absorption in the $n$th LL ($e+h$ absorption spectrum from 
   Fig.~\ref{fig5} is shown with gray bars for reference).
   Insets show magnified regions around the lowest two peaks 
   ``00'' and ``11''.
   Trion peaks ($X^-$) are identified.}
\label{fig8}
\end{figure}

Fig.~\ref{fig8} presents the $2e+h$ absorption spectra 
$\Omega({\cal E})$, calculated assuming that an electron
in the initial state is either in the lowest LL (ground 
state) or in a higher, $n=1$ LL (due to thermal excitation).
To observe superposition of many closely spaced small peaks,
each discrete $e\rightarrow2e+h$ transition was broadened
with a gaussian of width $\delta=0.5$~meV (main frames)
or 0.2~meV (insets).
As a reference, the ${\rm vac}\rightarrow e+h$ excitonic 
spectrum is shown with gray bars. 

The main result is that when the (great number of) main ODOS 
peaks ``$n'/nn$'' are shifted by $E_\psi=n'\hbar\omega_{c,e}$ 
to convert $\Omega(E)$ into $\Omega({\cal E})$, they all
fall exactly onto absorption spectrum of a bare exciton.
Presence of an additional electron does not cause shifting
or splitting of these main absorption lines, and it has 
insignificant effect on their relative intensities.
This demonstrates that, somewhat surprisingly, bare excitonic 
absorption is unaffected by a dilute electron gas (neither 
by renormalization of energy nor by transfer of intensity 
between LLs).
This result is obtained for a realistic quantum well, with
significant electron--hole asymmetry and LL mixing.

The effect of free electrons is emergence of additional 
(compared to the bare exciton) features in absorption spectrum, 
the strongest of them associated with the formation of trions.
In Fig.~\ref{fig8}, trion absorption peaks can be seen most 
clearly in the insets, in which the vicinities of peaks
``00'' and ``11'' have been magnified.
In reality, their intensity relative to the excitonic peaks
will depend on the filling factor and can be much higher 
than in our model (which represents a very dilute system
with only one free electron per $g=2Q+1$ states of the 
lowest LL).
Remarkably, the spin of strong trion-related features 
correlates with the parity of $n'-n$: singlet ($S=0$) peaks
appear for $n'=0$ below ``00'' and for $n'=1$ below ``11'',
while triplets ($S=1$) occur for the opposite combinations 
of $n'=0$ with ``11'' and $n'=1$ with ``00''.

\section{Experiment}

\begin{figure}
\includegraphics[width=3.2in]{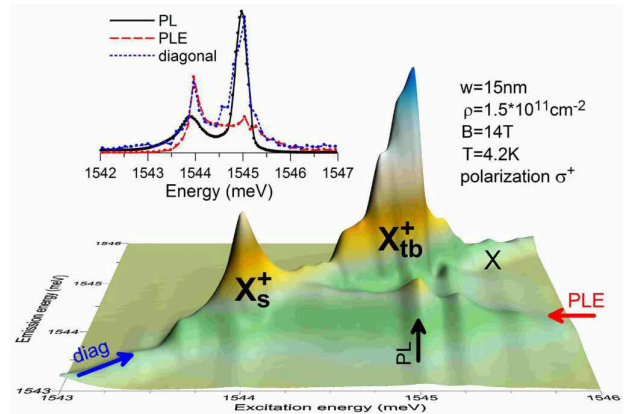}
\caption{(color online) 
    Polarized photoluminescence-excitation (PLE) 
   spectrum measured in a symmetric $w=15$~nm GaAs quantum well 
   with hole concentration $\varrho=1.5\cdot10^{11}$~cm$^{-2}$, 
   in magnetic field $B=14$~T, at temperature $T=4.2$~K.}
\label{fig9}
\end{figure}

To verify some of presented calculations we performed 
polarization-resolved photoluminescence (PL) and 
photoluminescence-excitation (PLE) experiments on a symmetric 
$w=15$~nm GaAs/AlAs quantum well, containing a valence holes 
gas with concentration $\varrho=1.5\cdot10^{11}$~cm$^{-2}$ 
and subject to a strong magnetic field.
The PLE experiment consisted of measuring polarization-resolved 
PL (emission intensity $I$ as a function of emission energy 
${\cal E}_{\rm out}$) for a series of excitation energies 
${\cal E}_{\rm in}$.
The spectra presented in Fig.~\ref{fig9} were recorded at $T=4.2$~K, 
in Faraday configuration, at $B=14$~T ($\nu\approx0.45$), and for 
the circular $\sigma^+$ polarization of light (corresponding to 
optical transitions of an electron in the excited spin state).
Emission intensity $I$ is plotted as a function of both 
${\cal E}_{\rm in}$ and ${\cal E}_{\rm out}$.
In the inset we showed the following three cross-sections.
(i) PL (emission) spectrum is $I({\cal E}_{\rm out})$ for
a fixed, high ${\cal E}_{\rm in}$.
Although it is typically measured for a much higher ${\cal 
E}_{\rm in}$, here we plot the data corresponding to resonant
excitation of $X^+_{\rm tb}$.
(ii) PLE spectrum is $I({\cal E}_{\rm in})$ for a fixed, low 
${\cal E}_{\rm out}$, here corresponding to the $X^+_{\rm s}$
recombination.
For regular dependence of relaxation on ${\cal E}_{\rm in}$, 
PLE is an indirect measure of absorption.
(iii) Diagonal cross-section $I({\cal E}_{\rm out}={\cal 
E}_{\rm in})$, usually difficult to detect due to strong
reflection of the incident light from the surface.
By smoothing and appropriate orientation of the surface 
with respect to the incident/reflected direction we were 
able to minimize this effect and record meaningful diagonal 
spectra.

In Fig.~\ref{fig9}, two strong peaks correspond to two optically 
active trions, $X^+_{\rm s}$ and $X^+_{\rm tb}$.
Other, minor features are artifacts (note a sizable 0.12~meV 
step in ${\cal E}_{\rm in}$).
The suppressed intensity of the $X$ (whose position relative to 
the $X^+$'s is anticipated from the following Fig.~\ref{fig10})
relative to the bright $X^+$'s results from a rather large hole 
concentration.
On the other hand, the strongest $X^+_{\rm tb}$ peak along the 
diagonal confirms earlier theoretical prediction for the trion 
oscillator strengths.\cite{Wojs00}

\begin{figure}
\includegraphics[width=3.2in]{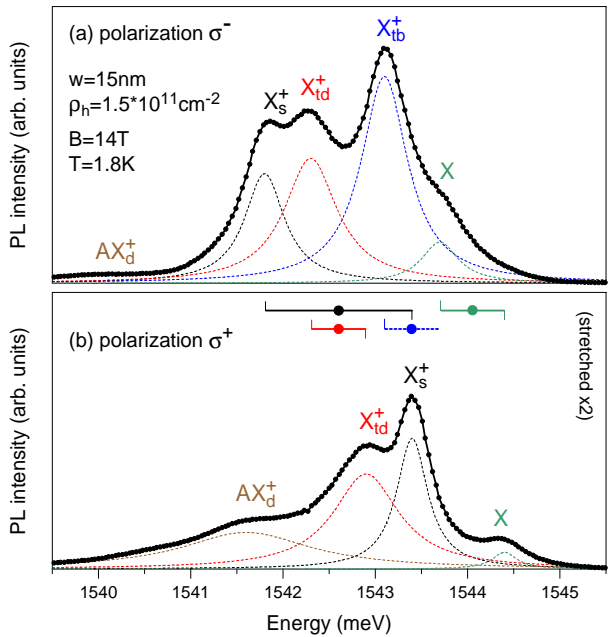}
\caption{(color online) 
   Polarized photoluminescence (PL) spectra measured in a 
   symmetric $w=15$~nm GaAs quantum well with hole concentration 
   $\varrho=1.5\cdot10^{11}$~cm$^{-2}$, in magnetic field $B=14$~T, 
   at temperature $T=1.8$~K.}
\label{fig10}
\end{figure}

In Fig.~\ref{fig10}, we plot a pair of polarized $\sigma^-$ and 
$\sigma^+$ PL spectra $I({\cal E}_{\rm out})$.
They were recorded in the same sample and at the same magnetic 
field as PLE of Fig.~\ref{fig9}, but for a higher excitation 
energy ${\cal E}_{\rm in}=2.57$~eV (wavelength 488~nm) and at 
low temperature $T=1.8$~K.
Three trion states ($X^+_{\rm s}$, $X^+_{\rm td}$, and 
$X^+_{\rm tb}$) along with the exciton are identified for the 
$\sigma^-$ polarization.
In the $\sigma^+$ spectrum, the $X^+_{\rm td}$ peak is weakened
and $X^+_{\rm tb}$ disappears completely due to spin polarization.
Additional low-energy peaks marked as $AX^+_{\rm d}=A^0+X^+$ show 
recombination of a spin doublet ($S={1\over2}$) ground state of 
a trion bound to an neutral acceptor located inside the well (cf.\ 
Fig.~\ref{fig4}); such impurity-bound trions are discussed 
elsewhere.\cite{tbp}

The unambiguous assignment of the peaks was possible from the 
analysis of the field evolution of the spectra, from $B=0$ 
beyond $B=14$~T, presented in a separate publication.\cite{tbp}
As noted by Glasberg {\em et al.},\cite{Glasberg99} the Zeeman 
splitting of different $X$ and $X^+$ states is very different 
(because of the wavevector dependence of the Land\'e $g$-factor 
for the holes).
We marked these splittings with color horizontal bars in 
Fig.~\ref{fig10}(b).
The large value of 1.6~meV for $X^+_{\rm s}$ (and $AX^+_{\rm d}$) 
compared to only 0.7~meV for $X$ and 0.6~meV for $X^+_{\rm td}$ 
is related to the occupation of a strongly repulsive 
zero-angular-momentum hole pair state, characteristic of the 
singlet trion (and of doublet $AX^+$).
The analysis of the pair-correlation function shows that this 
pair state is not occupied also in $X^+_{\rm tb}$ (despite triplet 
spin configuration).
This lets us expect similar $X^+_{\rm tb}$ and $X^+_{\rm td}$ 
Zeeman splittings, even though $X^+_{\rm tb}$ is only detected 
in one polarization.

Knowing the difference between $X$ and $X^+$ Zeeman splittings is 
necessary for a meaningful comparison of the trion binding energies 
$\Delta$ with the calculation of Fig.~\ref{fig2}(b) in which the 
Zeeman energy was ignored.
The Coulomb binding energies $\Delta$ are extracted from the 
experimental PL spectra by comparing the average $X$ and $X^+$ 
energies measured in both polarizations,\cite{Glasberg99} in 
Fig.~\ref{fig10}(b) marked by dots on the Zeeman bars.
In this way, we find $\Delta^+_{\rm s}=\Delta^+_{\rm td}\approx
1.4$~meV and $\Delta^+_{\rm tb}\approx0.6$~meV.
Compared to these values, the $x=1$ calculation of Fig.~\ref{fig2}(b)
predicting $\Delta=1.2$, 1.35, and 0.4~meV, respectively, slightly 
underestimates the binding of all three states.
The slight discrepancy could result from including only two lowest 
quantum well subbands and five lowest LLs, assuming equidistant LLs 
also for the holes, and ignoring the light-hole/heavy-hole mixing, 
all likely to enhance binding. 
The numerical tests indicate that neither higher subbands ($s\ge2$) 
nor the variation of LL spacing (beyond $n=1$) play a role, but the
the $n\le4$ restriction indeed has a noticeable ($\le0.2$~meV) effect 
on $\Delta$.

\begin{figure}
\includegraphics[width=3.2in]{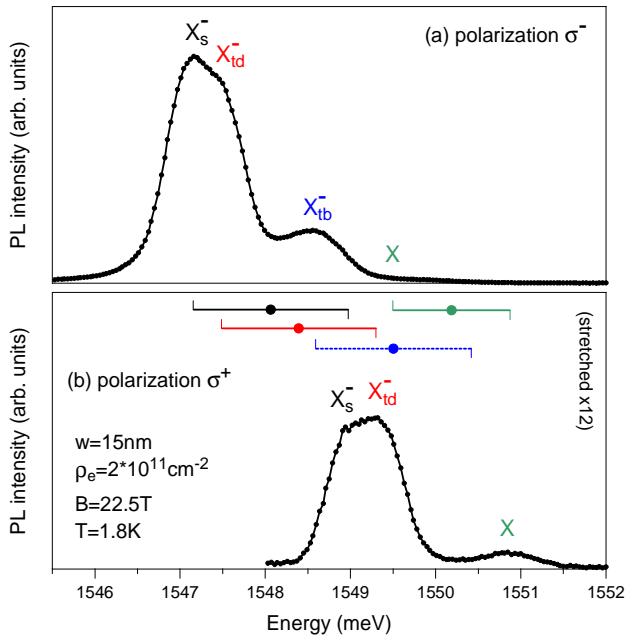}
\caption{(color online) 
   Polarized photoluminescence (PL) spectra measured in a 
   symmetric $w=15$~nm GaAs quantum well with electron concentration 
   $\varrho=2\cdot10^{11}$~cm$^{-2}$, in magnetic field $B=22.5$~T, 
   at temperature $T=1.8$~K.}
\label{fig11}
\end{figure}

The spectra in Fig.~\ref{fig11} are analogous to Fig.~\ref{fig10},
but they were recorded on an electron gas and involve negative trions.
The sample is also symmetric $w=15$~nm GaAs quantum well, but the
concentration is slightly higher, $\varrho=2\cdot10^{11}$~cm$^{-2}$,
so we had to use a stronger field, $B=22.5$~T to keep a sufficiently
low filling factor, $\nu\approx0.4$.
In both polarizations, the detected trions correspond to those of 
Fig.~\ref{fig10}.
The weak $X$ peak marked for $\sigma^-$ can be identified more 
convincingly when intensity is plotted in the logarithmic scale.
The Zeeman splittings found in this system are 1.8~meV for 
$X^-_{\rm s}$ and $X^-_{\rm td}$ (the same value was hence assumed
for $X^-_{\rm tb}$), and 1.4~meV for $X$.
Weaker variation of the splitting supports its attribution to the 
occupation of two-hole states.
Using these splittings, we find the following Coulomb trion binding 
energies: $\Delta^-_{\rm s}\approx2.1$~meV, $\Delta^-_{\rm td}\approx
1.8$~meV, and $\Delta^-_{\rm tb}\approx0.65$~meV.
Again, calculation of Fig.~\ref{fig2}(a) predicting $\Delta=2$, 1.4,
and 0.3~meV, respectively, slightly underestimates all these values.

\section{Conclusion}

We have carried out exact numerical diagonalization of realistic 
$e+h$ and $2e+h$ hamiltonians (including spin, Coulomb interactions, 
and mixing of LLs and quantum well subbands) on a Haldane sphere.
The parameters used for illustration are adequate for a symmetric 
15~nm GaAs quantum well in a magnetic field $B=10$~T.

Calculation of trion binding energies has been considerably 
advanced by taking the actual subband wavefunctions for the 
integration of Coulomb matrix elements and by including five 
LLs and two quantum well subbands for each electron and hole 
in the exact diagonalization.

From the full energy spectra we have calculated the density of 
states (DOS).
The main difference between DOS of $e+h$ and $2e+h$, representative
for the excitons with and without the presence of additional free 
electrons, is the emergence of discrete bound trion states below 
the excitonic tails and redistribution of DOS away from the LL 
peaks and filling the gaps between them.
The effect of an impurity on the $e+h$ DOS is also studied
as a function of its charge and position in the well.

For the full spectra of eigenstates, we have calculated the optical
oscillator strength $\Omega$.
The optical density of states (ODOS) of $2e+h$ shows no obvious
correlation with DOS.
It is fairly complicated, with a great number of strong transitions 
labeled by LL indices of the recombining $e$--$h$ pair and of the 
left-over electron.
However, the $e\rightarrow2e+h$ absorption spectrum is far simpler.
It is dominated by a series of LL peaks for the purely excitonic 
transitions ${\rm vac}\rightarrow e+h$.
These main peaks are neither shifted in energy, nor is their 
intensity noticeably suppressed or enhanced.
The presence of (and interaction with) an additional electron 
shows in form of additional weaker peaks.
Some of them are attributed to bound trion states (in the lowest 
and higher LLs).

The numerical results have been successfully compared with PL/PLE 
experiments carried out on a 2D hole and electron gases.
In particular, absorption and emission of the whole family of
both negative and positive trions in the lowest LL has been 
observed.

\begin{acknowledgments}

AW thanks J. J. Quinn for helpful discussions.
Work supported by grants: 
1P03B03230 and N20210431/0771 (Polish MNiSW), 
and RITA-CT-2003-505474 (EC).

\end{acknowledgments}


\begin{thebibliography}{99}

\bibitem{Lampert58}
M. A. Lampert,
   Phys. Rev. Lett. {\bf1}, 450 (1958).

\bibitem{Kheng93}
K. Kheng, R. T. Cox, Y. Merle d'Aubigne, F. Bassani, K. Saminadayar, 
and S. Tatarenko,
   Phys. Rev. Lett. {\bf71}, 1752 (1993).

\bibitem{Bassani75} 
F. Bassani and G. P. Parravicini, 
{\sl Electronic States and Optical Transitions in Solids}
(Pergamon, New York, 1975);
E. L. Ivchenko,
{\sl Optical Spectroscopy of Semiconductor Nanostructures}
(Alpha Science International Ltd., Oxford, 2005);
P. Hawrylak and M. Potemski,
   Phys. Rev. B {\bf56}, 12386 (1997).

\bibitem{Narvaez01}
G. A. Narvaez, P. Hawrylak, and J. A. Brum,
   Physica E {\bf9}, 716 (2001).

\bibitem{Huant90}
S. Huant, S. P. Najda, and B. Etienne,
   Phys. Rev. Lett. {\bf65}, 1486 (1990).

\bibitem{Dzyubenko02}
A. B. Dzyubenko,
   Phys. Rev. B {\bf65} 035318 (2002).

\bibitem{Stebe89}
B. Stebe and A. Ainane, 
   Superlatt. Microstruct. {\bf5}, 545 (1989).

\bibitem{Buhmann95}
H. Buhmann, L. Mansouri, J. Wang, P. H. Beton, N. Mori, 
L. Eaves, M. Henini, and M. Potemski,
   Phys. Rev. B {\bf51}, 7969 (1995).

\bibitem{Finkelstein95}
G. Finkelstein, H. Shtrikman, and I. Bar-Joseph,
   Phys. Rev. Lett. {\bf74}, 976 (1995);
   Phys. Rev. B {\bf53}, R1709 (1996).

\bibitem{Shields95a}
A. J. Shields, M. Pepper, M. Y. Simmons, and D. A. Ritchie,
   Phys. Rev. B {\bf52}, 7841 (1995).

\bibitem{Gekhtman96}
D. Gekhtman, E. Cohen, A. Ron, and L. N. Pfeiffer,
   Phys. Rev. B {\bf54}, 10320 (1996).

\bibitem{Astakhov99}
G. V. Astakhov, D. R. Yakovlev, V. P. Kochereshko, W. Ossau,
J. N\"urnberger, W. Faschinger, and G. Landwehr,
   Phys. Rev. B {\bf60}, R8485 (1999);
G. V. Astakhov, D. R. Yakovlev, V. P. Kochereshko, W. Ossau, 
W. Faschinger, J. Puls, F. Henneberger, S. A. Crooker, 
Q. McCulloch, D. Wolverson, N. A. Gippius, and A. Waag,
   {\em ibid.} {\bf65}, 165335 (2002).

\bibitem{Homburg00}
O. Homburg, K. Sebald, P. Michler, J. Gutowski, 
H. Wenisch, and D. Hommel,
   Phys. Rev. B {\bf62}, 7413 (2000).

\bibitem{Shields95b}
A. J. Shields, J. L. Osborne, M. Y. Simmons, M. Pepper, 
and D. A. Ritchie,
   Phys. Rev. B {\bf52}, R5523 (1995).

\bibitem{Glasberg99}
S. Glasberg, G. Finkelstein, H. Shtrikman, and I. Bar-Joseph,
   Phys. Rev. B {\bf59}, R10425 (1999).

\bibitem{Gorkov67}
L. P. Gor'kov and I. E. Dzyaloshinskii,
   Zh. Eksp. Teor. Fiz. {\bf53}, 717 (1967)
   [Sov. Phys.---JETP {\bf26}, 449 (1968)].

\bibitem{Bychkov81}
Yu. A. Bychkov, S. V. Iordanskii, and G. M. Eliashberg,
   Pis'ma Zh. Eksp. Teor. Fiz. {\bf33}, 152 (1981)
   [Sov. Phys.---JETP Lett. {\bf33}, 143 (1981)].

\bibitem{Kallin84}
C. Kallin and B. I. Halperin,
   Phys. Rev. B {\bf30}, 5655 (1984).

\bibitem{Lozovik02} 
Y. E. Lozovik, I. V. Ovchinnikov, S. Y. Volkov, L. V. Butov, 
and D. S. Chemla,
   Phys. Rev. B {\bf65}, 235304 (2002).

\bibitem{BarJoseph05}
I. Bar-Joseph,
   Semicond. Sci. Technol. {\bf20}, R29 (2005).

\bibitem{Peeters01}
F. M. Peeters, C. Riva, and K. Varga,
   Physica B {\bf300}, 139 (2001).

\bibitem{Wojs95} 
A. W\'ojs and P. Hawrylak, 
   Phys. Rev. B {\bf51}, 10880 (1995).

\bibitem{Palacios96}
J. J. Palacios, D. Yoshioka, and A. H. MacDonald,
   Phys. Rev. B {\bf54}, 2296 (1996).

\bibitem{Whittaker97}
D. M. Whittaker and A. J. Shields,
   Phys. Rev. B {\bf56}, 15185 (1997).

\bibitem{Hayne99}
M. Hayne, C. L. Jones, R. Bogaerts, C. Riva, A. Usher, F. M. Peeters, 
F. Herlach, V. V. Moshchalkov, and M. Henini,
   Phys. Rev. B {\bf59}, 2927 (1999).

\bibitem{Munteanu00}
F. M. Munteanu, Y. Kim, C. H. Perry, D. G. Rickel, J. A. Simmons, 
and J. L. Reno, 
   Phys. Rev. B {\bf61}, 4731 (2000);
F. M. Munteanu, D. G. Rickel, C. H. Perry, Y. Kim, J. A. Simmons, 
and J. L. Reno,
   {\em ibid.} {\bf62}, 16835 (2000).

\bibitem{Yusa01}
G. Yusa, H. Shtrikman, and I. Bar-Joseph,
   Phys. Rev. Lett. {\bf87}, 216402 (2001).

\bibitem{Vanhoucke01} 
T. Vanhoucke, M. Hayne, M. Henini, and V. V. Moshchalkov,
   Phys. Rev. B {\bf63}, 125331 (2001);
   {\em ibid.} {\bf65}, 041307 (2002);
   {\em ibid.} {\bf65}, 233305 (2002);
M. Hayne, T. Vanhoucke, and V. V. Moshchalkov,
   {\em ibid.} {\bf68}, 035322 (2003).

\bibitem{Astakhov05} 
G. V. Astakhov, D. R. Yakovlev, V. V. Rudenkov, P. C. M. Christianen, 
T. Barrick, S. A. Crooker, A. B. Dzyubenko, W. Ossau, J. C. Maan, 
G. Karczewski, and T. Wojtowicz
   Phys. Rev. B {\bf71}, 201312 (2005).

\bibitem{Wojs00}
A. W\'ojs, J. J. Quinn, and P. Hawrylak,
   Phys. Rev. B {\bf62}, 4630 (2000).

\bibitem{Riva00}
C. Riva, F. M. Peeters, and K. Varga,
   Phys. Rev. B {\bf61}, 13873 (2000);
   {\em ibid.} {\bf63}, 115302 (2001);
   {\em ibid.} {\bf64}, 235301 (2001).

\bibitem{Andronikov05} 
D. Andronikov, V. Kochereshko, A. Platonov, T. Barrick, 
S. A. Crooker, and G. Karczewski
   Phys. Rev. B {\bf72}, 165339 (2005).

\bibitem{Avron78} 
J. E. Avron, I. W. Herbst, and B. Simon,
   Ann. Phys. (N.Y.) {\bf114}, 431 (1978).

\bibitem{Dzyubenko00}
A. B. Dzyubenko and A. Y. Sivachenko,
   Phys. Rev. Lett. {\bf84}, 4429 (2000);
A. B. Dzyubenko,
   Solid State Commun. {\bf113}, 683 (2000).

\bibitem{Lerner81} 
I. V. Lerner and Yu. E. Lozovik,
   Zh. Eksp. Teor. Fiz. {\bf80}, 1488 (1981)
   [Sov. Phys. JETP {\bf53}, 763 (1981)].

\bibitem{Dzyubenko83} 
A. B. Dzyubenko and Yu. E. Lozovik,
   Fiz. Tverd. Tela {\bf25}, 1519 (1983)
   [Sov. Phys. Solid State {\bf25}, 874 (1983)].

\bibitem{Macdonald90} 
A. H. MacDonald and E. H. Rezayi,
   Phys. Rev. B {\bf42}, 3224 (1990).

\bibitem{Schuller02} 
C. Sch\"uller, K.-B. Broocks, Ch. Heyn, and D. Heitmann,
   Phys. Rev. B {\bf65}, 081301 (2002).

\bibitem{Jeukens02}
C. R. L. P. N. Jeukens, P. C. M. Christianen, J. C. Maan,
D. R. Yakovlev, W. Ossau, V. P. Kochereshko, T. Wojtowicz, 
G. Karczewski, and J. Kossut,
   Phys. Rev. B {\bf66}, 235318 (2002).

\bibitem{Plochocka04}
P. Plochocka, P. Kossacki, W. Maslana, J. Cibert, S. Tatarenko, 
C. Radzewicz, and J. A. Gaj,
   Phys. Rev. Lett. {\bf92}, 177402 (2004).

\bibitem{Stebe98}
B. Stebe, E. Feddi, A. Ainane, and F. Dujardin,
   Phys. Rev. B {\bf58}, 9926 (1998).

\bibitem{Laughlin83} 
R. B. Laughlin, 
   Phys. Rev. Lett. {\bf50}, 1395 (1983).

\bibitem{Wojs99}
A. W\'ojs, P. Hawrylak, and J. J. Quinn,
   Phys. Rev. B {\bf60}, 11661 (1999);
A. W\'ojs, I. Szlufarska, K.-S. Yi, and J. J. Quinn,
   {\em ibid.} {\bf60}, R11273 (1999).

\bibitem{Sanvitto02}
D. Sanvitto, D. M. Whittaker, A. J. Shields, M. Y. Simmons, 
D. A. Ritchie, and M. Pepper,
   Phys. Rev. Lett. {\bf89}, 246805 (2002).

\bibitem{Wu76} 
T. T. Wu and C. N. Yang,
   Nucl. Phys. B {\bf107}, 365 (1976).

\bibitem{Haldane83} 
F. D. M. Haldane, 
   Phys. Rev. Lett. {\bf51}, 605 (1983).

\bibitem{Cole97}
B. E. Cole, J. M. Chamberlain, M. Henini, T. Cheng, W. Batty, 
A. Wittlin, J. A. A. J. Perenboom, A. Ardavan, A. Polisski, 
and J. Singleton,
   Phys. Rev. B {\bf55}, 2503 (1997).

\bibitem{Tan90}
I.-H. Tan, G. L. Snider, L. D. Chang, and E. L. Hu,
   J. Appl. Phys. {\bf68}, 4071 (1990).

\bibitem{Finkelstein96}
G. Finkelstein, H. Shtrikman, and I. Bar-Joseph,
   Phys. Rev. B {\bf53}, 12593 (1996).

\bibitem{Dzyubenko04}
A. B. Dzyubenko,
   Phys. Rev. B {\bf69}, 115332 (2004).

\bibitem{tbp}
 A. W\'ojs, L. Bryja, J. Misiewicz, M. Potemski, D. Reuter, 
and A. Wieck,
   Proc. of XXXV Int. School on Physics of Semicond. Compounds 
   ``Jaszowiec 2006,'' Ustro\'n, Poland, June 17-23, 
   2006 (to appear in Acta Phys. Polon. A).

\end{thebibliography}
\end{document}